# The point spread function of electrons in a magnetic field, and the decay of the free neutron

D. Dubbers[1]*, L. Raffelt[1], B. Märkisch[1], F. Friedl[1]**, H. Abele[2]

[1]*Physikalisches Institut der Universität, Im Neuenheimer Feld 226, 69120 Heidelberg, Germany*

[2]*Atominstitut, Technische Universität, Stadionallee 2, 1020 Vienna, Austria*

*Corresponding author. *E-mail address*: dubbers@physi.uni-heidelberg.de

**Now at Institut für Umweltphysik, Universität Heidelberg

ABSTRACT

Experiments in nuclear and particle physics often use magnetic fields to guide charged reaction products to a detector. Due to their gyration in the guide field, the particles hit the detector within an area that can be considerably larger than the diameter of the source where the particles are produced. This blurring of the image of the particle source on the detector surface is described by a suitable point spread function (PSF), which is defined as the image of a point source. We derive simple analytical expressions for such magnetic PSFs, valid for any angular distribution of the emitted particles that can be developed in Legendre polynomials. We investigate this rather general problem in the context of neutron beta decay spectrometers and study the effect of limited detector size on measured neutron decay correlation parameters. To our surprise, insufficient detector size does not affect much the accuracy of such measurements, even for rather large radii of gyration. This finding can considerably simplify the layout of the respective spectrometers.

## 1. Introduction

In nuclear spectroscopy magnetic fields are frequently used to guide charged particles onto their detectors. This technique permits the use of relatively small detectors for a full $2\pi$ solid angle of detection, and even for $4\pi$ solid angle if two detectors are installed on either side of the source. In particle physics similar arrangements are found in time projection chambers. Such an arrangement is schematically shown in Fig. 1. The two detectors are then magnetically coupled not only to the source, but also to each other, which allows detecting particles in one detector that were backscattered on the other detector.

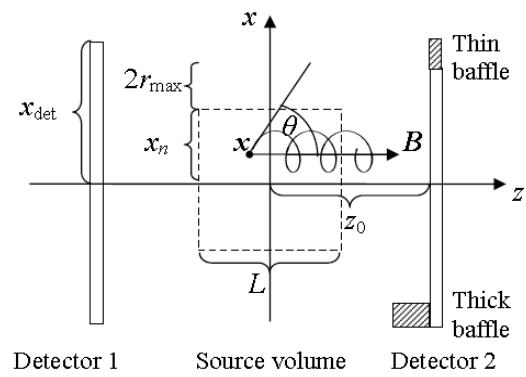

**Fig. 1.** A magnetic field $\boldsymbol{B}$ guides charged particles from a source volume of length $L$ and width $2x_n$ to square detectors of width $2x_{det}$. The helical trajectories of the particles with helix angle $\theta$ have diameters of up to $2r_{max}$. Hence the particles reach the detector over a square area of width $2x_n + 4r_{max}$. For the baffles see Sect. 3.4.

In particular, most modern spectrometers for the study of neutron decay parameters use magnetic fields to guide the decay electrons and/or protons onto their detectors. Examples are experiments or projects on neutron decay correlations, performed either with cold neutrons "in-beam" (abBA/PANDA [1], aCORN [2], aSPECT [3], Nab [4], PERC [5], Perkeo [6,7], Perkino [8], the latter with an artificial $\beta$-source, and Petersburg [9]) or performed with stored ultracold neutrons (UCNA [10,11], UCNB and UCNb [12]), as well as experiments on the neutron lifetime, in-beam (NIST [13]) or with stored UCN (HOPE [14], Mainz [15], PENeLOPE [16], UCNτ [17,18]). They all aim at relative accuracies of $10^{-3}$ or even $10^{-4}$. In favorable cases measurements can be entirely free of detector edge effects, provided the detectors are sufficiently large to intercept particles up to the largest radii of gyration. The data produced are used in various fields of nuclear and astrophysics, as well as for sensitive tests of the standard model of particle physics, for details see the reviews [19-21].

In addition to higher count rates, the magnetic coupling of the source to the detectors has another advantage that had already been pointed out in Lee and Yang's seminal paper [22] on the parity violation in weak interactions. In experiments on the weak decay of polarized nuclei, a magnetic guide field permits a clean cut between those particles emitted into one half space under angles $\theta < \pi/2$, with respect to the local field direction, and those emitted into the other half space under $\theta > \pi/2$. This makes the measurement of correlation coefficients like the $\beta$-asymmetry independent of the precise direction of the local magnetic field and of the precise positioning of the detectors.

Evidently, the particle detectors must be large enough to accept all events that contribute significantly to the signal. To be safe one would tend to add, all around the image of the source on the detector of width $2x_n$, an area that allows detecting all incoming decay particles up to their maximum diameter of gyration $2r_{max}$. The detector then must have a width of $2x_{det} \geq 2x_n + 4r_{max}$, as indicated in Fig. 1, and similarly for the height $2y_{det}$ of the detector. If some fraction of these particles is backscattered on one detector, a new helical trajectory is started that may require an additional safety margin of $2r_{max}$ around the area of detection.

In Sect. 2 we first derive the distribution of charged particles, originating from a point source, on the surface of the detector. In imaging theory such a distribution is called a *point spread function* or PSF. In our case we regard a beam-optical imaging system, consisting of a point source at $x = 0$, and a uniform magnetic guide field along $z$ that projects the charged particles onto the detectors installed at distances $\pm z_0$. We shall call the particle distribution function on the detector surface the *magnetic PSF*. We derive these magnetic PSFs analytically, and find surprisingly simple results for isotropic emission, for parity-violating asymmetric emission, and more generally for anisotropic charged particle emission of Legendre type, with the central result given in Eq. (25). There exist also analytical PSFs for the more general associated Legendre polynomials, as well as for the case of electron transport in non-uniform $B$-fields, but we postpone their discussion to a forthcoming publication.

In Sect. 3 we apply these PSFs to finite source volumes with position dependent source strength, and use the results for the ongoing neutron decay experiment PerkeoIII where $2r_{max}$ and $x_n$ are of similar magnitude. In particular, we are interested to know how insufficient detector size will influence the results on neutron decay parameters.

The study of this problem seems to be a rather elementary exercise. However, as we shall see, this investigation requires some care, the problem can be solved only using some subtle but very precise approximations, with a result that is both unexpected and comforting. At the same time we want to demonstrate how far one can go analytically before starting Monte Carlo simulations, which are difficult when investigating $10^{-4}$ effects for varying geometries. Charged particle guidance in a magnetic field is a widespread technique not only in nuclear physics, hence our results for the magnetic point spread functions may be of interest to a wider community.

## 2. The magnetic point spread functions

To be specific, we derive the magnetic point spread functions for the case of electron emission, although the results are valid for all kinds of charged particles. As an electron source we choose, again without loss of generality, free neutrons, instable against $\beta$-decay.

### 2.1 Source brightness

The exponential decay of a number of $N_n$ neutrons in an active decay volume generates a flux of electrons

$$\Phi_{e0} = -\dot{N}_n = N_n / \tau_n, \tag{1}$$

with the neutron lifetime $\tau_n = 880$ s. A volume element $d^3x$ located at position $x$ gives rise to the local electron flux element $\phi_{e0}(x) = d\Phi_{e0}(x) = dN_n(x)/\tau_n$, with neutron density $\rho_n(x)$, or

$$\phi_{e0}(x) = \rho_n(x) d^3x / \tau_n. \tag{2}$$

The subscript "0" indicates that the flux is that at the source. The local brightness of the electron source then is defined as

$$b_{e0}(x) = \frac{\partial^2 \phi_{e0}(x)}{\partial E \, \partial \Omega}, \tag{3}$$

with electron kinetic energy $E$ and solid angle $\Omega$ of electron emission.

### 2.2 Electron trajectories

Let the electrons be guided by the $B$-field towards one of the two detectors. For a uniform field the guiding center (Fig. 2) is a straight line along $z$. For emission under polar angle $\theta$ with respect to the field direction $z$ (Fig. 1), the radius of gyration of the electrons' helical trajectory is

$$r(E,\theta) = r_0(E)\sin\theta, \qquad (4)$$

with $0 \leq \theta \leq \pi$ and

$$r_0(E) = p/eB = \sqrt{E(E+2m)}/eB, \qquad (5)$$

with electron charge $e$, mass $m$, and momentum $p$, and with light velocity $c \equiv 1$, see for instance Reference [23].

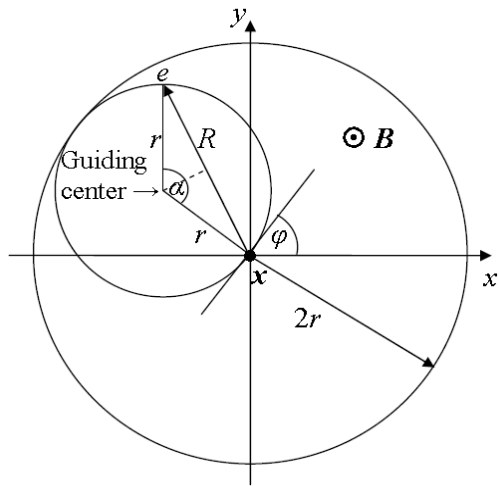

**Fig. 2.** An electron $e$, emitted in a uniform field $B$ from a point source at position $x$, under azimuth angle $\varphi$ and polar or helix angle $\theta$ (the latter shown in Fig. 1), moves with gyration radius $r$ along a helical path about the $B$-field. The axis of the helix is along $z$ (at right angles to the paper plane) and is called the guiding center. After spiralling through total phase angle $\alpha$, the electron reaches the detector surface a distance $R$ away from the "projected" source position $x$. The circle of radius $2r$ indicates the reach of electrons with varying $\varphi$.

From the source to the detector, for each complete cycle, the electron progresses along $z$ by the pitch of the helix of size

$$d = 2\pi r_0 \cos\theta. \qquad (6)$$

With increasing helix angle $\theta$, the pitch shortens and the diameter $2r$ of the helix widens, as shown in Fig. 3, which displays the relation $d^2 + C^2 = (2\pi r_0)^2$ between pitch $d$ and circumference $C = 2\pi r$ of the helix.

Upon arrival of the electron at the position $z_0$ of the detector, the total number of cycles is

$$n' = z_0 / d \qquad (7)$$

(with $n'$ generally not an integer). The total phase angle $\alpha$ reached at $z_0$ is determined by the starting angle $\theta$ via

$$\alpha = 2\pi n' = z_0 / (r_0 \cos\theta). \qquad (8)$$

The smallest phase angle occurs for electron emission under $\theta = 0$ into direction $z$ and is

$$\alpha_0 = z_0 / r_0 \equiv 2\pi n'_0. \qquad (9)$$

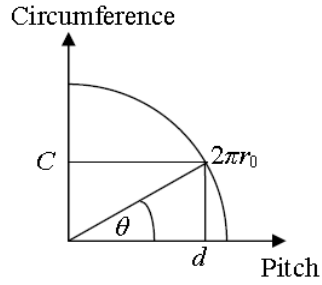

**Fig. 3.** Relation between pitch $d$, helix angle $\theta$, and circumference $C = 2\pi r$ of the helical trajectory of charged particles in a magnetic field.

In the experiments listed in the introduction, the maximum gyration radius $r_0$ is of order millimetres to centimeters, while the distance $z_0$ to the detector is of order meters, hence the minimum number of cycles from Eq. (9) is typically $n'_0 = z_0 / 2\pi r_0 \sim 100$. The emission angle with the largest statistical weight is at $\theta \approx \pi/2$ with $\cos\theta \to 0$, hence the typical number of cycles from Eq. (8) is $n' \gg 100$.

*2.3 The magnetic PSF on the detector*

A uniform magnetic field does not change the angular distribution of the particles on their way to the detectors, hence the imaged electron brightness $b_e$ arriving on the detector is the same as $b_{e0}$ at the source. An electron will arrive on the detector displaced from its projected starting point by a distance

$$R = 2r_0 \sin\theta \, |\sin\tfrac{1}{2}\alpha|, \qquad (10)$$

cf. Fig. 2. We define the magnetic point spread function for a point source of electrons of energy $E$ as the distribution function of these displacements $R$,

$$f_e(E,R)\,dR = \frac{1}{\phi'_0}\frac{\partial \phi'_e}{\partial R}\,dR, \qquad (11)$$

with the magnetically imaged spectral flux or "intensity"

$$\phi'_e = \partial \phi_e / \partial E. \qquad (12)$$

The energy dependence of $f_e$ comes in via this $\phi'_e$ and via $r_0(E)$ in Eq. (10). The infinitesimal increment $dR$ will be dropped in the following. The prefactor $1/\phi'_0$ is chosen such that $f_e = 1$ for $R = 0$ (assuming $f_e(E,R)$ non-divergent at $R = 0$). This prefactor also takes care of the positive sign of $f_e$. The quantity $2\pi R\, f_e(R)\,dR$ gives the probability for finding an electron at a distance between $R$ and $R+dR$ from its projected starting position. What is seen on the detector's $x$-$y$ surface is the rotationally symmetric function $f_e[(x^2+y^2)^{1/2}]$. Our first aim is to find the PSF of monoenergetic electrons with a given kinetic energy $E$.

We write the displacement $R$ in units of the helix diameter $2r_0$ as

$$\tilde{R} = R/2r_0. \qquad (13)$$

We use

$$\cos\theta = \alpha_0/\alpha = n'_0/n' \qquad (14)$$

from Eqs. (8) and (9), and express this "reduced displacement" $\tilde{R}$ as a function of the phase angle $\alpha$,

$$\tilde{R}(\alpha) = \sqrt{1 - \alpha_0^2/\alpha^2}\,|\sin \tfrac{1}{2}\alpha|, \qquad (15)$$

with $0 \le \tilde{R} \le 1$ and $\alpha \ge \alpha_0$. Fig. 4 shows $\tilde{R}(\alpha)$, for better visibility only for a smaller number of cycles from $n'_0 = 5.6$ to $n' = 20$.

Often the brightness at the source is some function of $\cos\theta$, frequently a Legendre polynomial $P_l(\cos\theta)$. With $b_e = \partial \phi'_e / \partial \cos\theta$ a function of $\cos\theta$ and $\tilde{R}$ a function of $\alpha$, the PSF Eq. (11) must be calculated as

$$f_e(E,\tilde{R}) = \frac{1}{\phi'_0}\frac{\partial \phi'_e}{\partial \cos\theta}\frac{\partial \cos\theta}{\partial \alpha}\frac{\partial \alpha}{\partial \tilde{R}}. \qquad (16)$$

The distribution $f_e$ is easily calculated as a function of phase angle, as described in the Appendix, and is shown in Fig. 5 for isotropic emission $\partial \phi'_e / \partial \cos\theta = 1$. In Fig. 4 for every half-integer value of $\alpha/2\pi$, the function $\tilde{R}(\alpha)$ has a maximum and is stationary with respect to $\alpha$, as can also be seen from Fig. 2, hence $\partial \alpha / \partial \tilde{R}$ and with it $f_e(\alpha)$ in Fig. 5 diverge at these values of $\alpha$.

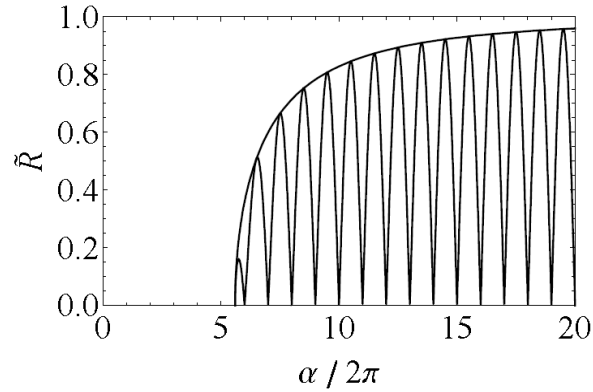

**Fig. 4.** Dependence of the reduced displacement $\tilde{R} = R/2r_0$ on the total phase angle $\alpha$. The envelope is given by the square root in Eq. (15). The function starts at the minimum number $n_0 = \alpha_0/2\pi$ of cycles for helix angle $\theta = 0$, given by Eq. (9), and continues to $\alpha \to \infty$ for $\theta \to \pi/2$. The envelope of this function is given by the square root in Eq. (15).

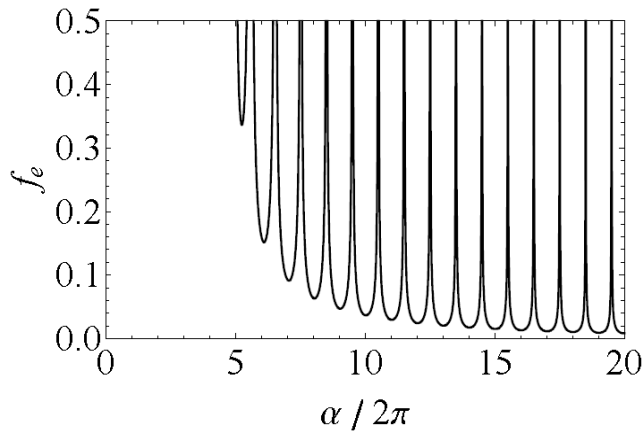

**Fig. 5.** Dependence of the electron distribution $f_e(\alpha)$ on the total phase angle $\alpha$. Singularities appear at every half-integer number of cycles. The challenge is to convert this function $f_e(\alpha)$ into a PSF $f_e(\tilde{R})$ by inverting $\tilde{R}(\alpha)$.

However, what we need is not the distribution $f_e(\alpha)$, but the PSF $f_e(\tilde{R})$, and the question is how to obtain the multivalued inverse function $\alpha(\tilde{R})$ from the function $\tilde{R}(\alpha)$ displayed in Fig. 4. To achieve this we exploit the fact that for all experiments under discussion, $\tilde{R}(\alpha)$ is the product of the extremely rapidly varying function $\sin\!½\alpha$ and the slowly varying envelope $\sin\theta(\alpha) = (1-\alpha_0^2/\alpha^2)^{1/2}$.

In view of the very large number of cycles it is an excellent approximation that between one cycle and the next, the change of the envelope is negligible (except possibly for the very first cycles near $n_0'$ where the envelope changes rapidly). For each $n^{th}$ cycle the envelope then assumes a constant value

$$s_n = (1 - n_0^2/n^2)^{1/2}, \tag{17}$$

with integer $n \geq n_0$. For a given number of cycles $n$, Eq. (15) can then be resolved for $\alpha$,

$$\alpha_n(\tilde{R}) = 2\pi n + 2\arcsin(\tilde{R}/s_n), \text{ with } \tilde{R} \leq s_n, \tag{18}$$

to be inserted for $\alpha$ into the $n^{th}$ partial PSF that we call $\hat{f}_{e,n}(E,\tilde{R})$, valid between $\alpha = 2\pi n$ and $\alpha = 2\pi(n+1)$, with the arcsine function limited to the first cycle of its argument.

The PSF as a function of $\tilde{R}$ then is

$$f_e(E,\tilde{R}) = \sum_{n=n_0}^{\infty} \hat{f}_{e,n}(E,\tilde{R}), \tag{19}$$

details of the derivation are given in the Appendix.

In practice, the infinite sum ends at some cutoff value $N$. Fig. 6 shows $f_e(E,\tilde{R})$ for $n_0 = 100$ (with summation starting at $n = n_0+1$), with a cutoff at $N = 10^4$. With $\cos\theta = n_0/N = 0.01$, this corresponds to a cutoff angle at $\theta = 89.5°$. Between $N = 3 \times 10^3$ with $\theta = 88°$ and $N = 10^4$, The singularities in Fig. 5 reappear as peaks in Fig 6, positioned at $\tilde{R} = s_n$. These wiggles disappear when measured in a detector with less extreme spatial resolution than shown in the figure. When these wiggles are neglected, the resulting $f_e(\tilde{R})$ appears to be constant up to the largest permitted displacements $\tilde{R} = R/2r_0 = 1$, that is, the displacements $R$ are evenly distributed over the whole allowed interval $0 \leq R \leq 2r_0$. The normalized *intensity PSF* for monoenergetic charged particles then is

$$f_e(E,R) = 1 \text{ for } R \leq 2r_0(E), \tag{20}$$

and $f_e = 0$ otherwise, as shown by the horizontal straight line in Fig. 6. This result will be rederived in Sect. 2.5 in a different approximation. It should however be kept in mind that the peaks visible in Fig. 6 are not artifacts, but should be measurable for not too high $n_0$ in a detector with sufficient spatial resolution. (Note added: a recent preprint [32] reports experimental detection of these wiggles.)

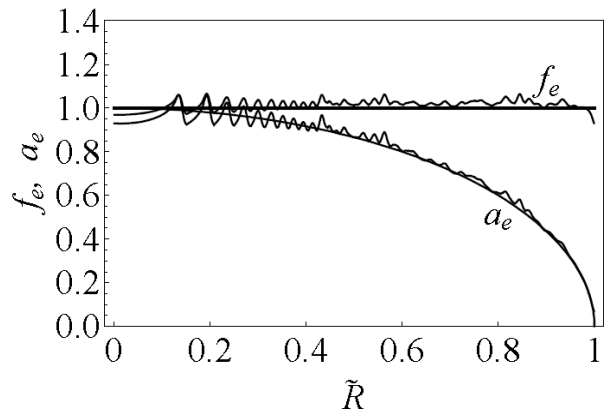

**Fig. 6.** The wiggling curves show the magnetic PSFs from the infinite sum method Eq. (19), as they would look when measured with sub-millimeter spatial resolution in a detector with a Gaussian width of $\sigma = 0.5$ % of the full PSF range $\tilde{R} = 1$. The curve labelled $f_e$ is the intensity PSF, the curve $a_e$ is the asymmetry PSF that will be discussed in Sect. 2.4. The smooth curves are from Eqs. (20) and (23), and will be derived in Sect- 2.5 under less stringent conditions.

The above result of a constant PSF $f_e(R)$ for isotropic emission is corroborated by a Monte Carlo study, with the two free parameters angle $\varphi$ (Fig. 2) and $\sin\theta$ (for isotropic emission) chosen at random. In the simulation we use the fact that each full cycle along the helix resembles the preceding cycle. Therefore only the phase angle $\alpha$ as counted from its last full cycle before arrival on the detector is of relevance for the size of $R$ on the detector. This phase angle then is $\alpha = 2\pi \,\text{Mod}(z_0, d)/d$, with pitch $d$ and detector position $z_0$, and depends on $\theta$ via $d$ from Eq. (6). The resulting PSF again is found constant, with $\chi^2 = 85$ for 99 degrees of freedom.

Our result of constant $f_e$ is qualitatively consistent with two measurements, however of very low statistics, presented in Ref. [24] in the upper part of their Fig. 6.29. For these two measurements the magnetic field is low enough that $R_{\text{max}} = 2r_0$ clearly exceeds the spatial resolution of the detector.

*2.4 Magnetic PSF for the electron asymmetry*

In the decay of polarized neutrons the electron brightness Eq. (3) is asymmetric with respect to emission angle $\theta$. When the neutron polarization is flipped between the two states $+P_n$ and $-P_n$, the brightness is flipped as well between $b_{e+}$ and $b_{e-}$,

$$b_{e\pm} \propto 1 \pm \beta(E) A P_n \cos\theta, \quad (21)$$

with the parity-violating asymmetry parameter $A$ and the electron velocity parameter $\beta = v_e/c$ (neglecting a tiny ~1% energy dependence of $A$). The asymmetry

$$\frac{b_{e+} - b_{e-}}{b_{e+} + b_{e-}} = \beta(E) A P_n \cos\theta \quad (22)$$

from a point source will then depend on the distance $R$ of electron impact on the detector, too. We therefore define also a point spread function for the asymmetry. To do so we have to insert $\partial \phi'_e / \partial \cos\theta = \beta(E) A P_n \cos\theta$ into Eq. (16). For monoenergetic electrons this amounts to multiplying each $f_{e,n}(\tilde{R})$ in Eq. (19) by $\cos\theta = n_0/n$ (setting $\beta A P_n$ to one), to obtain the corresponding asymmetry PSF $a_{e,n}(\tilde{R})$ (the usual division by $f_{e,n}(\tilde{R})$ is omitted, assuming from Fig. 6 that is unity). The sum over $n$ then gives the lower, slightly staggering curve $a_e(\tilde{R})$ in Fig. 6, calculated with the same limiting parameters $n_0$ and $N$ as $f_e(\tilde{R})$, again normalized to one at the origin. The limiting value for $\tilde{R} \to 1$ is only realized for electrons emitted near $\theta = \pi/2$ where the asymmetry Eq. (22) vanishes.

Fig. 6 also shows as a smooth line what we call the *asymmetry PSF*

$$a_e(E, R) = a_0 \sqrt{1 - (R/2r_0)^2}, \text{ for } R \leq 2r_0(E), \quad (23)$$

and $a_e = 0$ otherwise, with $a_0(E) = \beta(E) A P_n$. Equation (23) is a special case of the more general Eq. (25) below that we shall derive next.

*2.5 A more direct derivation of the PSFs*

In an exact description of charged particle gyration, to a given starting angle $\theta$ belongs one unique phase angle $\alpha$, both angles being linked by Eq. (8), and one unique displacement $R$ from Eq. (10). This holds for any azimuth angle $\varphi$ of particle emission. In the preceding section we chose the approximation that, within each cycle shown in Fig. 5, the phase angle $\alpha$ can be varied independent of the helix angle $\theta$, which latter was kept constant, and we found that this approximation was excellent for all practical purposes.

We now go one step further and assume that $\alpha$ and $\theta$ are completely independent variables, not only within one cycle of $\alpha$. This assumption is founded on Eq. (14), in which the correlation between $\cos\theta$ (the source of the smooth envelope of $\tilde{R}$), and $\alpha$ (the source of the rapid oscillations of $\tilde{R}$) is progressively lost with increasing number of cycles $n$, except possibly at the maxima $\tilde{R} = s_n$ where $\tilde{R}(\alpha)$ more closely follows the envelope function. This approach will allow us arriving directly at the PSFs given in Eqs (20) and (23) that are suggested by our infinite-sum results from Eq. (19). Furthermore, this approach will lead to analytic solutions for very general angular distributions of the emitted particles.

The validity of this second method is also supported by the finding that, as we shall see, it gives the same result as the first method of infinite sums (for not too small $n_0$, fulfilled in all experiments). As this first method is evidently an excellent approximation, the same should be true for the second method with its less stringent approximation. The details of this approach will be described in the Appendix.

When the displacement $\tilde{R}$ is fixed, both variables $\alpha$ and $\theta$ still remain linked to each other via Eq. (10). For a given $\tilde{R}$ we then have to integrate $f_e(\tilde{R})$ over all values of $\alpha$ compatible with Eq. (10). This integration, as discussed in the Appendix, leads to an analytic solution for all angular distributions of the type

$$\mathrm{d}\phi'_e \propto \cos^m\theta \tag{24}$$

namely, the *general anisotropy PSF* for monoenergetic electrons

$$f_e^{(m)}(E,R) = [1 - (R/2r_0)^2]^{m/2}, \text{ for } R \leq 2r_0(E), \tag{25}$$

and $f_e^{(m)} = 0$ otherwise. When we know the solution for arbitrary powers of $\cos\theta$, then we know the solution for any Legendre polynomial $P_l(\cos\theta)$.

For $m = 0$, Eq. (25) coincides with the intensity PSF $f_e$ from Eq. (20), and for $m = 1$ it coincides with the asymmetry PSF $a_e$ from Eq. (23). We compared Eq. (25) with the corresponding infinite-sum solutions from Eq. (19) for values up to $m = 8$, and found throughout agreement of similar quality as in Fig. 6, as long as the minimum number of cycles did not fall much below $n_0 \sim 10$.

The result Eq. (25) can be further generalized to angular distributions that are developed in associated Legendre polynomials, involving terms of type $\partial \phi'_e / \partial \cos\theta = \cos^m\theta \sin^{l-m}\theta$, with integers $m \leq l$. ", by integration of Eq. (43) in the Appendix. These analytical results, together with those found for electron transport in non-uniform fields, will be derived in a forthcoming publication.

To check our approximation of independence of the variables $\alpha$ and $\theta$ made above, another Monte Carlo was run with random input of all three independent parameters $\alpha$, $\sin\theta$, and $\varphi$. Again a constant $f_e$ was found, with $\chi^2 = 98$ for 99 degrees of freedom.

*2.6 PSFs for allowed β-emitters*

Next, instead of monoenergetic electrons, we take a point-like $\beta$-emitter with endpoint energy $E_0$ and an allowed electron energy spectrum

$$\phi'_\beta(E) = p(E+m)(E-E_0)^2, \tag{26}$$

with electron momentum $p = [E(E+2m)]^{1/2}$ as in Eq. (5).

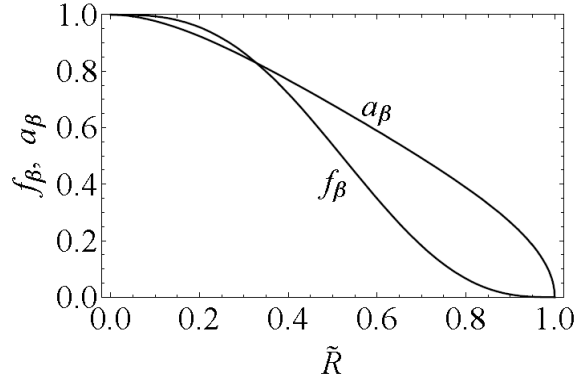

**Fig. 7.** Magnetic PSFs for an allowed $\beta$-spectrum Eq. (27). The lower curve $f_\beta(\tilde{R})$ is the $\beta$-intensity-PSF Eq. (27), the upper curve $a_\beta(\tilde{R})$ is the $\beta$-asymmetry-PSF Eq. (29).

For an allowed $\beta$-spectrum, the PSF of an isotropically emitting $\beta$-source then is

$$f_\beta(E_0, R) = \int_{E_{\min}}^{E_0} \phi'_\beta(E) f_e(E, R) \, dE. \tag{27}$$

The lower limit of integration, with $r_{\max} = r_0(E_0)$,

$$E_{\min} = \sqrt{E_0(E_0 + 2m)(R/2r_{\max})^2 + m^2} - m, \tag{28}$$

follows from the requirement $R < 2r_0$. For the isotropic case with $f_e = 1$, the integration of Eq. (27) has an algebraic result, shown in Fig. 7, though too long to be reproduced here.

For an allowed $\beta$-spectrum, the PSF of the energy-integrated $\beta$-asymmetry then is

$$a_\beta(E_0, R) = \frac{1}{f_\beta(E_0, R)} \int_{E_{\min}}^{E_0} \phi'_\beta(E) a_e(E, R) \, dE, \tag{29}$$

also shown in Fig. 7.

## 3. Applications in neutron decay

Neutron $\beta$-decay

$$n \to p^+ + e^- + \bar{\nu}_e \tag{30}$$

is characterized by a number of parameters that describe the angular correlations between the various measurable particle momenta and spins involved. These correlation coefficients, called *a*, *A*, *B*, *C*, *D*, *R*, etc., often lead to angular distributions of type Eq. (21). The $\beta$-asymmetry *A*, which describes the correlation between electron momentum and neutron spin, was mentioned in the previous section. In the following we shall also regard the proton and extend our treatment to the correlation between proton momentum and neutron spin, called the proton asymmetry *C*, on which new experiments are in preparation. As the sizes of proton momenta in neutron decay are comparable to electron momenta, the same is true for their gyration parameters in a magnetic field.

### 3.1 Neutron flux profiles

In experiments on neutron $\beta$-decay, the source is an ensemble of either ultracold neutrons stored in a trap, or of cold neutrons in a beam, with typical active source volumes of several liters. The electron flux element $\phi_\beta(x)$, or alternatively its energy spectrum $\phi'_\beta = \partial \phi_\beta / \partial E$, then must be integrated over the decay volume. For an in-trap experiment like UCNA, with trap height of order 1 dm such that gravitational effects can be neglected, UCN density $\rho_n(x)$, and with it the electron flux from Eq. (2), is rather uniform. For an in-beam experiment on a cold neutron guide like PerkeoIII, to which our magnetic PSFs will be applied in the following, $\rho_n(\mathbf{x})$ can be calculated with a simple formula that we shall derive next.

PerkeoIII uses a 6×6 cm$^2$ section of the cold neutron guide H113 [25] of the Institut Laue-Langevin (ILL) in Grenoble, which delivers a thermal equivalent flux density $\phi_n = 3 \times 10^9$ cm$^{-2}$s$^{-1}$ of nearly 100% polarized neutrons. PerkeoIII usually works with a monochromatic pulsed neutron beam, with de Broglie wavelength $\lambda_n = 0.56$ nm, 10 ms cycle time, on-off ratio 1:5, and 1.4×10$^7$ freely flying neutrons in each pulse. In its guide field of $B = 0.15$ T, pointing along the beam axis $z$, electrons and protons have gyration diameters of up to 5 cm, widening to 7 cm at the detectors where the field has dropped to $B = 0.09$ T. To register all decay particles, including those backscattered on one of the two scintillation detectors, the detector area would have to be quite large, with negative consequences for the detector's energy resolution, spatial uniformity, background rate, and for the size of the instrument. Therefore it is worthwhile to study in detail the measurement errors arising from insufficient detector size.

For a straight guide of rectangular cross section, the neutron flux density $\phi_n(\mathbf{x})$ is separable in the $x$ and $y$ coordinates. For a monochromatic neutron beam of velocity $\upsilon_n$, the same is true for the neutron density $\rho_n(\mathbf{x}) = \phi_n(\mathbf{x})/\upsilon_n$. Furthermore, a straight neutron guide emits neutrons isotropically, up to the critical angle $\theta_c$ of total neutron reflection, with $\theta_c/\lambda_n = 0.024$/nm for H113. Hence the local neutron density is proportional to the product of the opening angles $\theta_n$ and $\theta'_n$ along $x$ and $y$, under which the square neutron guide exit is seen from within the active neutron volume, or

$$\rho_n(\mathbf{x}) \propto \theta_n(x,z)\theta'_n(y,z). \tag{31}$$

This separation ansatz holds rather well also for the curved "ballistic supermirror" neutron guide H113, see Ref. [26].

When the neutrons are collimated by a system of rectangular apertures, the opening angles, and with it the neutron density, can be calculated for any given position $(x, z)$ or $(y, z)$ within the beam volume with a single line of code

$$\theta_n(x,z) = \mathrm{Min}(\theta_c, \theta_1^+, \theta_2^+, ...) - \mathrm{Max}(-\theta_c, \theta_1^-, \theta_2^-, ...), \tag{32}$$

where $\theta_i^\pm$ are the two limiting angles under which the $i$th aperture is seen in the ($x$-$z$)-plane, see Fig. 8, and similarly for $\theta'_n(x,y)$.

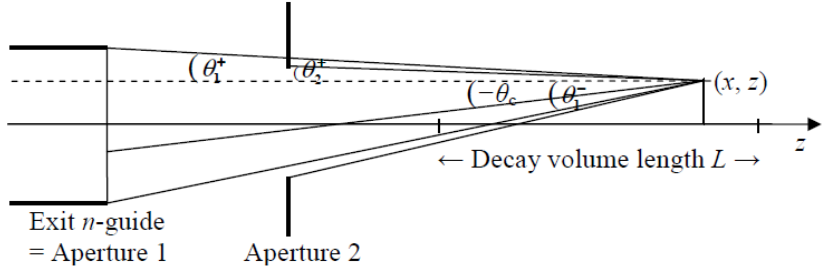

**Fig. 8.** Beam geometry for the calculation of neutron flux density profiles from Eq. (32), with the critical angle $\theta_c$ of total neutron reflection, and the angles $\theta_i^\pm$ under which the opening of the *i*th aperture is seen from the point $(x, z)$. In the example shown the opening angle Eq. (32) is $\theta_n(x,z) = \theta_2^+ - \theta_1^-$.

For PerkeoIII, the guide exit is seen, from within the neutron decay volume, through a collimator made up of four collinear apertures with square openings, each of area 6×6 cm². For a pulsed beam with $\lambda_n$ = 5.6 nm, which is the usual mode of operation, the length of the decay volume is typically $L = 2$ m, and its maximum width $2x_n = 12$ cm (for a continuous beam, $L = 3.8$ m and $2x_n = 23$ cm). The neutron flux along one approximately straight magnetic field line along $z$ must be added up numerically and is proportional to

$$\bar{\rho}_n(x) \propto \int \theta_n(x,z) \mathrm{d}z, \tag{33}$$

with $\theta_n(x, z)$ from Eq. (32), and similarly for $\bar{\rho}_n(y)$.

*3.2 Electron and proton flux profiles*

Folding of this neutron density profile $\bar{\rho}_n(x)\bar{\rho}_n(y)$ with the appropriate electron PSF $f_e$ or $f_\beta$ then gives the electron flux profiles. Folding, for instance, with the PSF $f_\beta$ gives the energy-integrated two-dimensional electron flux profile

$$\phi_\beta(x,y) \propto \iint f_\beta\left(\sqrt{(x-x')^2 + (y-y')^2}\right) \bar{\rho}_n(x')\bar{\rho}_n(y') \mathrm{d}x'\mathrm{d}y', \tag{34}$$

while folding with $f_e$ gives the energy resolved profile $\phi'_e(x,y) = \partial \phi_e(x,y)/\partial E$.

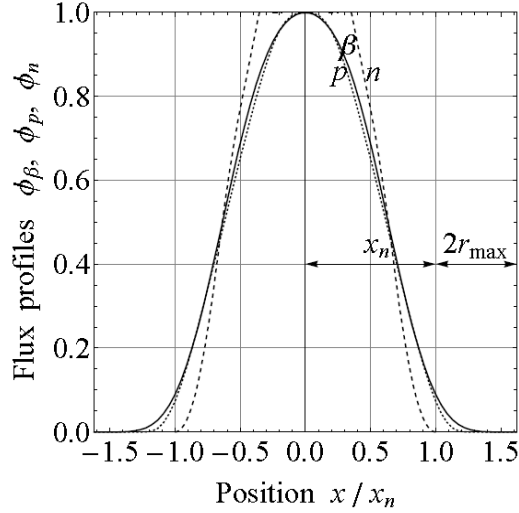

**Fig. 9.** Neutron, electron, and proton beam $x$-profiles for neutron decay in PerkeoIII (for $y = 0$). All distributions are set to one at the origin. Dashed line: neutrons $n$ at wavelength $\lambda_n = 0.56$ nm, from Eq. (33); full line: electrons $\beta$, from Eq. (34); dotted line: protons $p$, from numerical calculation. Position $x$ is given in units of $1/2 \times$ maximum width of the neutron beam, which is $x_n = 8$ cm. Although the initial maximum gyration diameter $2r_{max} \approx 5$ cm is quite large, the electron and proton distributions rather closely follow the neutron distribution.

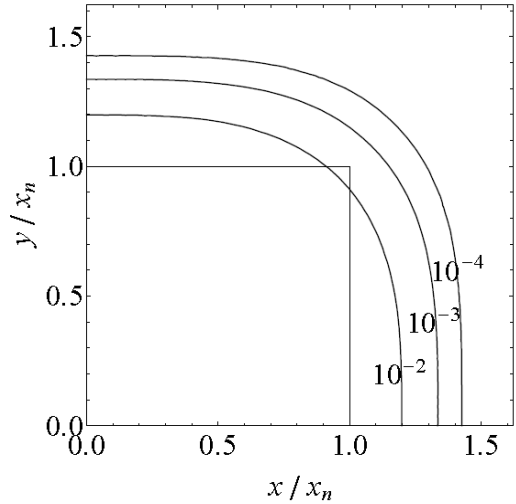

**Fig. 10.** Contour plot of the electron distribution $\log \phi_\beta(x, y)$ for neutron decay in PerkeoIII. The straight lines at $x = x_n$ and $y = x_n$ indicate the maximum lateral extension of the neutron decay volume. The contours indicate where electron intensity has dropped to $10^{-2}$, $10^{-3}$, and $10^{-4}$ of its central value at $x = y = 0$.

Figure 9 shows as solid line a cut of this electron distribution $\phi_\beta(x, 0)$ along $y = 0$. The dotted line gives the proton distributions, based on the proton spectra derived, for instance, in Ref. [27]. For comparison, the dashed line gives the neutron distribution $\rho_n(x, 0)$. The difference between the neutron and the electron or proton beam profiles is small, much smaller than the maximum electron or proton reach $2r_{max}$ might suggest.

The logarithmic contour plot Fig. 10 shows the $\beta$-distribution $\phi_\beta(x, y)$ over the detector surface. As already seen in Fig. 9, the electron intensity outside the neutron beam is largely negligible. To find the corresponding energy-integrated electron asymmetry profile, we simply replace $f_\beta$ in Eq. (34) by $a_\beta$, and divide this by the flux profile from Eq. (34),

$$A_\beta(x, y) \propto \frac{1}{\phi_\beta(x, y)} \iint a_\beta\left(\sqrt{(x-x')^2 + (y-y')^2}\right) \bar{\rho}_n(x') \bar{\rho}_n(y') \, dx' \, dy'. \tag{35}$$

*3.3 Losses due to insufficient detector size*

Our aim is to find out how the measured neutron decay parameters depend on the linear dimension of the square detector used in the experiment. To this end we integrate the electron intensity Eq. (34) over the detector area,

$$\Phi_\beta(x_{det}) = \int_{-x_{det}}^{x_{det}} \int_{-x_{det}}^{x_{det}} \phi_\beta(x, y) \, dx \, dy, \tag{36}$$

and vary both width = height $x_{det}$ of the detector, up to a maximum value $X_{det} = x_n + 2x_{max}$. The same then is done for the $\beta$-asymmetry Eq. (35), as well as for proton intensity and proton asymmetry.

Figure 11 shows, for both electrons and protons, the fraction of particles missing the detectors

$$\frac{\Delta\Phi_\beta}{\Phi_\beta} = 1 - \frac{\Phi_\beta(x_{det})}{\Phi_\beta(X_{det})}. \tag{37}$$

Also shown are the corrections for the $\beta$-asymmetry $\Delta A/A$ and for the proton asymmetry $\Delta C/C$ that must be applied when the detectors have not the full size needed to intercept all decay products.

To our surprise, even for a linear detector size not larger than the width of the active neutron beam volume, $x_{det}/x_n = 1.0$ in the figure, the correction on the $\beta$-asymmetry parameter $A = -0.1176(11)$ (lower full line) is only half the relative error $\pm 0.9 \times 10^{-2}$ of its present world average [28]. The same is true for the correction on the proton asymmetry $C = -0.2377(26)$ (lower dotted line), with a world average relative error of $\pm 1.2 \times 10^{-2}$. This means that detector area (and with it background, signal losses, resolution, etc.) can be reduced to less than half of the area needed to intercept all decay particles.

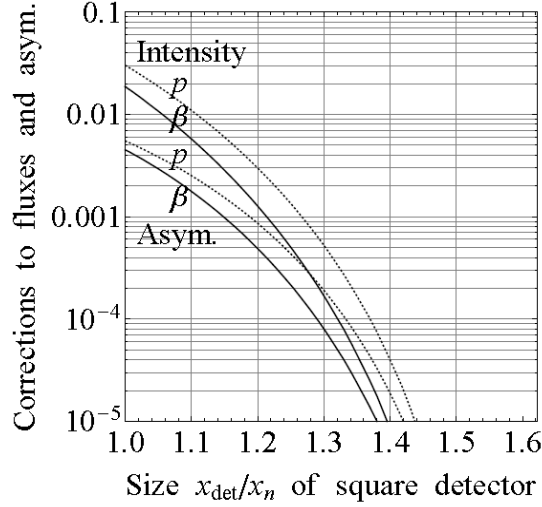

**Fig. 11.** Effects due to insufficient detector size: Corrections $\Delta X/X$, for total intensities $X = \Phi_\beta$, $\Phi_p$ and for decay asymmetries $X = A$, $C$, both for electrons ($\beta$) and protons ($p$), as functions of the width of the square detector. The plot starts at $x_{\text{det}} = x_n$, where the detector covers just the (projected) area of the neutron beam, and ends at $x_{\text{det}} = x_n + 2r_{\text{max}}$, where the detector has reached the size such that no corrections are needed (neglecting electron backscattering).

Figures 9 to 11 are calculated under the assumption of a uniform field between source and detectors. A non-uniform field as used in PerkeoIII will lead to small changes in these plots, which, however, will not alter the main conclusions of the present article, as will be discussed in a forthcoming publication.

*3.4 Further edge effects*

Our calculations incorporate already what is called a detector edge effect. Such edge effects make particle losses near the edges of the detector depend both on the particles' angle and energy, and thus affect the value of the measured correlation coefficients. However, other small effects related to the detector edge may come into play: In our derivations we have assumed that all particles hitting the detector surface at positions $x \leq x_{\text{det}}$ and $y \leq x_{\text{det}}$ are fully registered, while particles arriving at $x > x_{\text{det}}$ or $y > x_{\text{det}}$ completely miss the detector. For protons, with their very short ranges in the detector material, this assumption is realistic, but electrons with maximum ranges of several mm require further discussion. We distinguish the following cases.

(1) Those electrons that hit the detector at a few mm distance to its edges may leave the detector through one of its narrow lateral faces, so they do not deliver their full energy to the detector. However, for our example of neutron decay in PerkeoIII, we find that even for a detector as small as $x_{\text{det}} = x_n$, only a $10^{-3}$ fraction of all electrons arrive at the detector within this margin, and only another small fraction of these are scattered out laterally. The number of electrons concerned diminishes further when, to be safe, we choose a detector size of, say, $x_{\text{det}} = 1.1\, x_n$, cf. Fig. 11.

(2) Those electrons that miss the detector surface may, on their continuing helical path, still hit the detector on its lateral face and contribute to the signal. In PerkeoIII the radii of gyration are larger than the mm-range of electrons, therefore this effect involves more electrons than the aforementioned effect (1). The effect can be avoided by covering the detector's lateral face with an absorber. In the case of a scintillation detector as used in Perkeo, this absorber can be a light guide coupled laterally to the scintillator, forming the "thin baffle" shown in Fig. 1. For a solid state detector, the detector mounting may play this role. Electrons arriving on the surface of this absorber may be scattered into the active volume of the detector, but this occurs at a similarly small rate as for the inverse process (1) of scattering out of the detector material. – Hence, in view of the present accuracy goal of $10^{-4}$ for correlation coefficients in neutron decay, these additional edge effects can be neglected also for detectors of considerably reduced size.

(3) A "thick baffle", also shown in Fig. 1, has one surface lying parallel to the local magnetic field. For the case of a constant neutron density across this surface (as met in previous Perkeo experiments), the effect of this baffle on the $\beta$-asymmetry $A$ can be calculated analytically, see [29], and [30,31] for a refined analysis. A sufficiently thick baffle removes essentially all particles whose diameter of gyration is within the baffle's reach. In contrast, a bare detector as studied in this paper, or a detector equipped with a lateral thin baffle as in case (2), misses only a smaller subset of the particles stopped by the thick baffle, namely those whose helical path happens to cross the detector plane outside the detector's active area. Therefore for the cases studied in this article a thick baffle brings no advantage.

**Conclusions**

We have derived the point spread functions for charged particles in a uniform magnetic guide field, for monoenergetic particles and for allowed electron and proton spectra in nuclear $\beta$-decay, for isotropic particle emission and for more general angular distributions. For monoenergetic particles the central result for the PSFs is given in Eq. (25). We have applied these magnetic PSFs to angular correlation experiments in neutron decay. At the present level of accuracy, one can safely neglect the electrons or protons that gyrate outside the active neutron decay volume, even for rather low magnetic fields with large radii of gyration. For neutron decay instruments like PerkeoIII, this means that the size of the electron or proton detectors, and with it the size of the instrument, can be chosen rather small without compromising the results on the neutron decay parameters.

**Appendix**

We derive the approximate "infinite-sum" magnetic PSF Eq. (19). The exact function $f_e(\alpha)$ shown in Fig. 5 is derived from Eq. (16), using $\partial\cos\theta/\partial\alpha$ from Eq. (8) and $(\partial\tilde{R}/\partial\alpha)^{-1}$ from Eq. (15), with the result

$$f_e(\alpha) = \left| \frac{2\alpha_0(\alpha^2 - \alpha_0^2)}{\pi\alpha(\alpha^2 - \alpha_0^2)\cos\tfrac{1}{2}\alpha + \alpha_0^2\sin\tfrac{1}{2}\alpha} \right|, \tag{38}$$

From this we derive the infinite-sum solution (19), which was made und the assumption that, within one cycle of the gyrating electron, the envelope function (17) is independent of the phase angle $\alpha$. To this end we separate the phase angle $\alpha$ into its integer and fractional parts, see Eq. (18). With the substitution $n'/n'_0 = \alpha/\alpha_0$ from Eq. (14), with continuous $n'$, this reads $n' = n + (1/\pi)\arcsin(\tilde{R}/s_n)$. With $\cos(n\pi) = \pm 1$ and $\sin(n\pi) = 0$, we are left with $\cos\tfrac{1}{2}\alpha = \pm\sqrt{1 - \tilde{R}^2/s_n^2}$ and $\sin\tfrac{1}{2}\alpha = \pm\tilde{R}/s_n$, which we insert all into Eq.(38).

For $n \geq n_0 \gg 1$, valid for all practical purposes, we have $n \approx n'$, and the partial PSFs in the sum (19) become

$$\hat{f}_{e,n}(\tilde{R}) = \frac{2n_0(n^2 - n_0^2)}{\pi n(n^2 - n_0^2)\sqrt{n^2(1 - \tilde{R}^2) - n_0^2} + n\, n_0^2\, \tilde{R}} \quad \text{for } \tilde{R} \leq s_n, \tag{39}$$

that is, for integer number $n \geq n_0/\sqrt{1 - \tilde{R}^2}$, and $f_{e,n}(\tilde{R}) = 0$ otherwise.

Second, we derive the *general anisotropy PFS*, Eq. (25) in Sect. 2.4, which was made under the assumption that the phase angle $\alpha$ and the helix angle $\theta$ are independent parameters. For brevity, in the displacement $R$ from Eq. (10) we make the substitutions $u = \cos\theta$, $v = \sin\theta$, and rewrite the reduced displacement $\tilde{R} = R/2r_0$ as

$$\tilde{R} = v\sqrt{1-u^2}. \tag{40}$$

For a given displacement $\tilde{R}$, the independent angle $\alpha$ (modulo $2\pi$; hence $0 \leq v \leq 1$) is then allowed to vary from $\alpha_{min} = 2\arcsin(R/2r_0)$ to $\alpha_{max} = \pi$, as shown in Fig. 12.

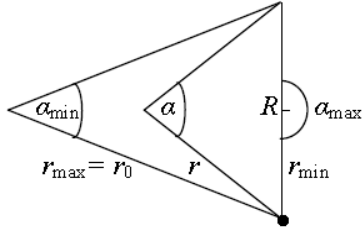

**Fig. 12.** When the electron's point of impact on the detector is fixed at a distance $R$ from the projected source position marked by the dot •, the permitted gyration radii $r = r_0\sin\theta$ vary from $r_{min} = R/2$ to $r_{max} = r_0$, hence $\sin\theta$ varies from a minimum value $R/2r_0$ to 1. At the same time, linked by Eq. (40), $\sin\frac{1}{2}\alpha$ varies from $R/2r_0$ to 1. Permitted are also the events that lie symmetrically to the right of the vertical line marked $R$.

With $\tilde{R}$ fixed, the PSF Eq. (16) then must be integrated over all allowed phase angles $\alpha$,

$$f_e(E,\tilde{R}) = \frac{2}{\phi'_0} \int_{\alpha_{min}}^{\pi} \frac{\partial \phi'_e}{\partial u} \frac{\partial u}{\partial \tilde{R}} d\alpha = \frac{2}{\phi'_0} \int_{\tilde{R}}^{1} \frac{\partial \phi'_e}{\partial u} \frac{\partial u}{\partial \tilde{R}} \frac{\partial \alpha}{\partial v} dv, \tag{41}$$

where the latter integral is introduced because it generally has analytic solutions. The additional factor of 2 takes care of the events that lie symmetrically to the right of the vertical line marked $R$ in Fig. 12. In contrast to the exact derivation of Eq. (38), the derivatives under the integral are obtained from $u(v) = (1 - \tilde{R}^2/v^2)^{1/2}$ and $v(\alpha) = \sin\frac{1}{2}\alpha$, or their inverse function, using Eq. (40), without recourse to Eq. (14).

The second integral in Eq. (41) then reads

$$f_e(E,\tilde{R}) = \frac{2}{\phi'_0} \int_{\tilde{R}}^{1} \frac{\partial \phi'_e}{\partial u} \frac{\tilde{R}}{v\sqrt{v^2 - \tilde{R}^2}\sqrt{1-v^2}} dv. \tag{42}$$

For isotropic particle emission with $\partial\phi'_e/\partial u = 1$, this PSF is simply a constant $\pi/\phi'_0$, independent of $\tilde{R}$, as was suggested already by the infinite-sum result for $f_e$ in Fig. 6. As a check we transform Eq. (42) into an integral over $u$. Using for the limits of integration the results from the figure caption of Fig. 4, we obtain the slightly simpler expression

$$f_e(E,\tilde{R}) = \frac{2}{\phi'_0} \int_0^{\sqrt{1-\tilde{R}^2}} \frac{\partial\phi'_e}{\partial u} \frac{\mathrm{d}u}{\sqrt{1-u^2-\tilde{R}^2}} \qquad (43)$$

where the integral, for isotropic emission, again gives $\pi/2$. For anisotropic angular distributions, when developed in Legendre polynomials that involve terms of type $\partial\phi'_e/\partial u = u^m = \cos^m\theta$, integration of Eq. (42) gives the general result Eq. (25).

### Acknowledgements

This work was supported by the Priority Programme SPP 1491 of the German DFG and the Austrian FWF.